\documentclass[prl,reprint,superscriptaddress,showpacs,showkeys,preprintnumbers,nofootinbib]{revtex4-1}
\usepackage{amsmath,amsfonts,amssymb,amsbsy}
\usepackage{mathrsfs,latexsym}
\usepackage{graphicx}
\usepackage{color,colortbl}
\definecolor{darkred}{rgb}{0.4,0.0,0.0}
\definecolor{darkgreen}{rgb}{0.0,0.4,0.0}
\definecolor{darkblue}{rgb}{0.0,0.0,0.4}
\usepackage[bookmarks,linktocpage,colorlinks,
    linkcolor = darkred,
    urlcolor  = darkblue,
    citecolor = darkgreen
    ]{hyperref}
\usepackage{macros_alpha}
\usepackage{dcolumn}
\usepackage{booktabs}
\newcolumntype{C}{>{$}c<{$}}
\AtBeginDocument{
\heavyrulewidth=.08em
\lightrulewidth=.05em
\cmidrulewidth=.03em
\belowrulesep=.65ex
\belowbottomsep=0pt
\aboverulesep=.4ex
\abovetopsep=0pt
\cmidrulesep=\doublerulesep
\cmidrulekern=.5em
\defaultaddspace=.5em
}
\usepackage{multirow}

\usepackage{defs}

\begin{document}
\pdfoutput=1

\newcommand{\LUV}{\mu_{\rm cut}}
\newcommand{\mupt}{\mu_{\rm PT}}
\newcommand{\muhadi}[1]{\mu_{\rm had #1}}
\newcommand{\muref}{\mu_{\rm ref}}

\preprint{CERN-TH-2017-129}
\preprint{DESY 17-088}
\preprint{WUB/17-03}
\title{The strong coupling from a nonperturbative determination of\\
the $\Lambda$ parameter in three-flavor QCD
}
\date{\today}
\collaboration{\href{https://www-zeuthen.desy.de/alpha/}{ALPHA collaboration}}

\author{Mattia~Bruno}
\affiliation{Physics Department, Brookhaven National Laboratory, Upton, NY 11973, USA}
\author{Mattia~Dalla~Brida}
\affiliation{Dipartimento di Fisica, Universit\`a di Milano-Bicocca and \\
INFN, Sezione di Milano-Bicocca, Piazza della Scienza 3, 20126 Milano, Italy}
\author{Patrick~Fritzsch} \affiliation{Theoretical Physics Department, CERN, 1211 Geneva 23, Switzerland}
\author{Tomasz~Korzec} \affiliation{Department of Physics, Bergische Universit\"at Wuppertal, Gau{\ss}str. 20,
42119 Wuppertal, Germany}\author{Alberto~Ramos} 
\affiliation{Theoretical Physics Department, CERN, 1211 Geneva 23, Switzerland}
\author{Stefan~Schaefer}
\affiliation{John von Neumann Institute for Computing (NIC), DESY, Platanenallee~6, 15738~Zeuthen, Germany}
\author{Hubert~Simma}
\affiliation{John von Neumann Institute for Computing (NIC), DESY, Platanenallee~6, 15738~Zeuthen, Germany}
\author{Stefan~Sint}\affiliation{School of Mathematics and Hamilton Mathematics Institute, Trinity College Dublin, Dublin 2, Ireland}\author{Rainer~Sommer}
\affiliation{John von Neumann Institute for Computing (NIC), DESY, Platanenallee~6, 15738~Zeuthen, Germany}
\affiliation{Institut~f\"ur~Physik, Humboldt-Universit\"at~zu~Berlin, Newtonstr.~15, 12489~Berlin, Germany}

\begin{abstract}
We present a lattice determination of the $\Lambda$ parameter in three-flavor
QCD and the strong coupling at the Z pole mass.  Computing the nonperturbative
running of the coupling in the range from $0.2\,\GeV$ to $70\,\GeV$, and using
experimental input values for the masses and decay constants of the pion
and the kaon, we obtain $\Lambda_{\overline{\rm MS}}^{(3)}=341(12)\,\MeV$.
The nonperturbative running up to very high energies guarantees that systematic
effects associated with perturbation theory are well under control.  Using the
four-loop prediction for 
$\Lambda_{\overline{\rm MS}}^{(5)}/\Lambda_{\overline{\rm MS}}^{(3)}$  yields
$\alpha^{(5)}_{\overline{\rm MS}}(m_{\rm Z}) = 0.11852(84)$.
\end{abstract}

\maketitle

\section{Introduction}

An essential input for theory predictions of high energy processes, in
particular for phenomenology at the LHC
\cite{Dittmaier:2012vm,Heinemeyer:2013tqa,Accardi:2016ndt,deFlorian:2016spz}, is the QCD coupling
$\alpha_s(\mu)=\gbar^2_s(\mu)/(4\pi)$ at energy scales $\mu \sim m_\mathrm{Z}$
and higher. In this work we present a sub-percent determination of the strong
coupling at the Z pole mass using the masses and decay constants of the pion
and kaon as experimental input and lattice QCD as computational tool.

Perturbation theory (PT) predicts the energy dependence of the coupling as
\begin{equation}
    \gbar_s^2(\mu) \simas{\mu\to\infty} 
    \frac{1}{2 b_0 \log(\mu/\Lambda_s) + (b_1/b_0) \log\log(\mu/\Lambda_s)}
    + \ldots
    \label{e:asy}
\end{equation}
in terms of known positive coefficients, $b_{0,1}$, and a single parameter,
$\Lambda_s$, which can also 
serve as the nonperturbative scale of the theory.
The label $s$, called scheme, summarizes all details of the exact definition of
$\gbar_s$.  Conventionally one chooses the so-called $s=\msbar$
scheme~\cite{Bardeen:1978yd}, but $\Lambda$ parameters in different schemes can
be exactly related with a one-loop computation~\cite{Celmaster:1979km}.

Our computation of $\alpha_\msbar$ is based on a determination of the
three-flavor $\Lambda$ parameter.  To outline the steps of our determination,
we write
\begin{equation}
  \Lambda^{(3)}_\msbar =
  \frac{\Lambda^{(3)}_\msbar}{\mupt}
  \times \frac{\mupt}{\muhadi{}}
  \times \frac{\muhadi{}}{\fpik}
  \times \fpik^{\rm PDG}\,.
  \label{e:master}
\end{equation}
As experimental input we use the PDG values~\cite{Olive:2016xmw} for the
following combination of decay constants 
\begin{equation}
  \fpik \equiv \frac13 (2\fk +\fpi) = 147.6\,\MeV\,.
  \label{e:fpikpdg}
\end{equation}
The key elements are then the determination of the ratio of scales
$\mupt/\muhadi{}$ and the ratio $\muhadi{}/\fpik$, i.e., our hadronic scale in
units of $\fpik$.  Both computations are performed in a fully nonperturbative
way.

By choosing a large enough scale $\mupt$ and including higher orders of PT in
\eqref{e:asy}, the ratio $\Lambda^{(3)}_\msbar/\mupt$ can be determined with
negligible errors.

With $\nf>2$ flavors, so far a single work \cite{Aoki:2009tf} contains such a
computation with all steps, including the connection of low energy $\muhadi{}$
to large $\mupt$, using numerical simulations and a step scaling strategy.
This strategy, developed by the ALPHA collaboration
\cite{Luscher:1991wu,Luscher:1993gh,deDivitiis:1994yz,Jansen:1995ck}, supresses
the systematic errors from the use of PT. 

Here, we put together (and briefly review) the first factor in
\eq{e:master} and our recent significant improvements in statistical and
systematic precision in the second
one~\cite{Brida:2016flw,DallaBrida:2016kgh}, and finally add the missing
third one.

QCD with $\nf=3$  is the phenomenologically relevant effective theory at
energies $E < m_\mathrm{charm}$ with small~\cite{Bruno:2014ufa,Korzec:2016eko}
corrections of order $(E/m_\mathrm{charm})^2$.  However, for theory predictions
of high energy processes, with $E \sim m_\mathrm{Z}$ and higher, the five- and
six-flavor theories are needed. Fortunately, the ratios
$\Lambda^{(\nf)}_\msbar/\Lambda^{(3)}_\msbar,\;\nf=4,5,6$ are known to very high 
order in PT, and successive order contributions decrease rapidly.  This
enables us to convert our $\Lambda^{(3)}_\msbar$ to precise values for
$\alpha^{(5)}_\msbar(m_\mathrm{Z})$ and $\alpha^{(6)}_\msbar(\mulhc)$, which
can be used for high energy phenomenology.  Further below, we will critically
discuss the use of PT in this step.

\section{Strategy} 
\begin{table}
  \caption{\label{tab:scales}
           Summary of various scales used in this work.
          }
  \def\arraystretch{1.1}
  \begin{ruledtabular}
  \begin{tabular}{lll}
    Scale definition                                         & purpose                                           & $\mu/\GeV$ \\\hline
    \multirow{2}{*}{$\mupt = 16\, \mu_{0}$}                  & \scalebox{0.9}{matching with the asymptotic     } & \multirow{2}{*}{$\approx 70$}\\[-0.2em]
                                                             & \scalebox{0.9}{perturbative behavior            } & \\[0.1em]
    \multirow{2}{*}{$\gbar_{\rm SF}^2(\mu_{0})=2.012$}       & \scalebox{0.9}{nonperturbative matching         } & \multirow{2}{*}{$\approx 4$} \\[-0.2em] 
                                                             & \scalebox{0.9}{between the GF and SF schemes    } & \\[0.1em]
    \multirow{2}{*}{$\gbar_\infty^2(\muref^\star)=1.6\pi^2$} & \scalebox{0.9}{setting scale in physical units  } & \multirow{2}{*}{$\approx 0.5$}\\[-0.2em] 
                                                             & \scalebox{0.9}{by experimental value for $\fpik$} & \\[0.1em]
    \multirow{2}{*}{$\gbar_{\rm GF}^2(\muhadi{})=11.31$}     & \scalebox{0.9}{matching between GF scheme       } & \multirow{2}{*}{$\approx 0.2$}\\[-0.2em] 
                                                             & \scalebox{0.9}{and infinite-volume scheme       } & \\
  \end{tabular}
  \end{ruledtabular}
\end{table}
A nonperturbative definition of a coupling is easily given.  Take a
short-distance QCD observable, depending on fields concentrated within a 4-d
region of Euclidean space of linear size $R=1/\mu$ and with a perturbative expansion 
\bes
 \obs_s(\mu) = k\,\gbar^2_\msbar(\mu)\,[1 + c_1^s\gbar^2_\msbar(\mu)+\ldots] 
   \label{e:obsexp} \,.
\ees
Then the nonperturbative  coupling, 
\begin{eqnarray}
  \gbar^2_s(\mu) \equiv \obs_s(\mu) / k =
   \gbar^2_\msbar(\mu) + c_1^s\gbar^4_\msbar(\mu) +\ldots\,,
   \label{e:gengdef}
\end{eqnarray}
runs with $\mu$. This property also allows us to define scales $\mu$ by fixing
$\gbar_s^2(\mu)$ to particular values (see Table~\ref{tab:scales}). However,
there is a challenge to reach the asymptotic region of small
$\gbar_s^2(\mu)$, where \eq{e:asy} is useful and its corrections can be
controlled, using lattice simulations. 

\subsection{Challenge}

\begin{figure}[t]
  \includegraphics[width=\columnwidth]{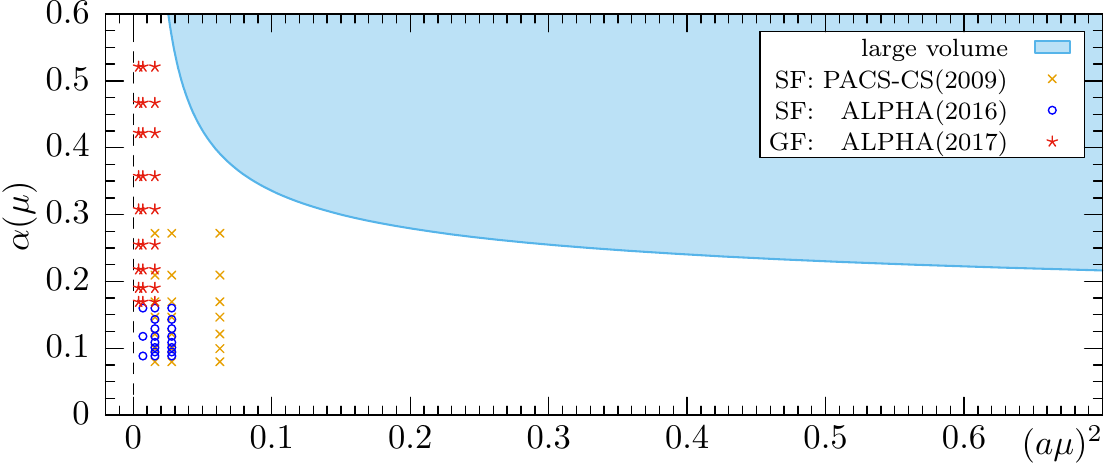} 
  \caption{\label{f:landscape}
           The shaded area shows the $a > 0.04\,\fm$ region of large volume
           results which dominate the present PDG and FLAG
           estimates~\cite{Olive:2016xmw,Aoki:2016frl} of $\alpha_\msbar(\mu)$
           (curve evaluated to two loops for $\nf=3$ with $\Lambda^{(3)}_\msbar
           =341\,\MeV$). The data points on the left are finite-size scaling
           computations~\cite{Aoki:2009tf,Brida:2016flw,DallaBrida:2016kgh}. 
          }
\end{figure}

Numerical computations involve both a discretization length, the lattice
spacing $a$, and a total size of the system $L$, that is simulated. For
standard observables, control over finite volume effects of order $\exp(-\mpi
L)$ requires $L$ to be several fm. At the same time, one needs to suppress
discretization errors and should extrapolate $(a\mu)^2 \to 0$ at fixed
$\mu$. The necessary restrictions
\bes
  L \gg 1/\mpi\,,\quad 1/a \gg \mu \quad \Rightarrow \; L/a \ggg \mu/\mpi
  \label{e:chall}
\ees 
translate into very large lattices. \Fig{f:landscape} displays the region in
$\alpha(\mu)$ vs. $(a\mu)^2 $ for the range $a\geq 0.04\,\fm$ which can be
realized nowadays in large volumes ($\mpi L\geq 4$).  This shaded region is
quite far from small coupling and small $(a\mu)^2$.

\subsection{Finite-size schemes}

The way out has long been known~\cite{Luscher:1991wu,Jansen:1995ck}.  One may
identify $R=L=1/\mu$ by choosing $\obs_s$ to depend only on the scale $L$, not
on any other ones.  Finite-size effects become part of the observable rather
than one of its uncertainties.  \Eq{e:chall} is then relaxed to
\bes
  L/a\gg 1 \,,
\ees
such that $L/a=10 - 50$ is sufficient. 

Different scales $\mu$ are then connected by the step scaling function 
\begin{equation}
  \sigma(u) \equiv \bar g_s^2(\mu/2) \Big|_{\bar{g}_s^2(\mu)=u}\,.
\end{equation}
It describes scale changes by discrete factors of two, 
in contrast to the $\beta$-function which is defined by infinitesimal changes.
For a chosen value of $u$, $\sigma(u)$ can be computed by determining $\gbar^2$
on lattices of size $L/a$ and $2L/a$ and performing an extrapolation
$a\rightarrow 0$ at $L=1/\mu$, fixed through $\gbar^2(\mu)=u$.  In fact, in the
process also the $\beta$-function can be computed as long as $\sigma(u)$ is a
smooth function of $u$. A recent detailed description of step scaling is given
in \cite{Sommer:2015kza}.

\section{Running coupling in the three-flavor theory between $200\,\MeV$ and $100\,\GeV$} 
We impose Schr\"odinger Functional (SF) boundary conditions on all
fields~\cite{Luscher:1992an,Sint:1993un}, i.e.,     Dirichlet boundary
conditions in Euclidean time at $x_0=0, L$, and periodic boundary conditions in
space with period $L$. 
With this choice, one can define different renormalized couplings in the
massless theory~\cite{Luscher:1992an,Fritzsch:2013je,DallaBrida:2016kgh} 
and complications with perturbation theory~\cite{GonzalezArroyo:1981vw} are
avoided.

First, we consider the SF coupling~\cite{Luscher:1992an,Sint:2012ae},
$\bar{g}_{\rm SF}(\mu)$, which measures how the system reacts to a particular
change of the boundary conditions. When computed by Monte Carlo methods, this
coupling  has a statistical uncertainty that scales as $\Delta_{\rm
stat}\gbar_{\rm SF}^2 \sim \gbar_{\rm SF}^4$, leading to good
 precision at high
energies. Moreover, its $\beta$-function is known to NNLO~\cite{Bode:1998hd,Bode:1999sm}.
These two properties make it an ideal choice to match with the asymptotic
perturbative regime of QCD. 

Second, one can use the gradient flow (GF) to define renormalized
couplings~\cite{Luscher:2010iy}. 
The flow field $B_\mu(t,x)$ is  the solution of the gradient flow equation
\begin{equation}
  \label{eq:flow}
  \begin{split}
    \partial_t B_\mu(t,x) &= 
    D_\nu G_{\nu\mu}(t,x)\,, \qquad     \\
    G_{\mu\nu}(t,x) &= \partial_\mu B_\nu - \partial_\nu B_\mu
    + [B_\mu,B_\nu]\,,
  \end{split}
\end{equation}
with the initial value $B_\mu(0,x)$ given by the original gauge field.  In
infinite volume a renormalized coupling is defined by
\begin{equation}
  \label{eq:t0}
  \gbar_{\infty}^2(\mu) =
  \frac{16\pi^2}{3}\times
  t^2\langle E(t) \rangle
  \Big|_{\mu=1/\sqrt{8t}} \;,
\end{equation}
using the action density at positive flow time~\cite{Luscher:2010iy},
$E(t)=\tfrac{1}{4}G_{\mu\nu}^a(t,x)G_{\mu\nu}^a(t,x)$. In finite volume 
the coupling $\gbar^2_{\rm GF}(\mu)$ is
defined  by imposing a fixed relation,  $\sqrt{8t} = cL$, between the flow time
and the volume~\cite{Fodor:2012td,Fritzsch:2013je}. Details can be found in the original
work~\cite{DallaBrida:2016kgh}.   Since the statistical precision is generally
good and scales as $\Delta_{\rm stat}\gbar_{\rm GF}^2 \sim \gbar_{\rm GF}^2$,
this coupling is well suited at low energies.  

\begin{figure}
  \includegraphics[width=\columnwidth]{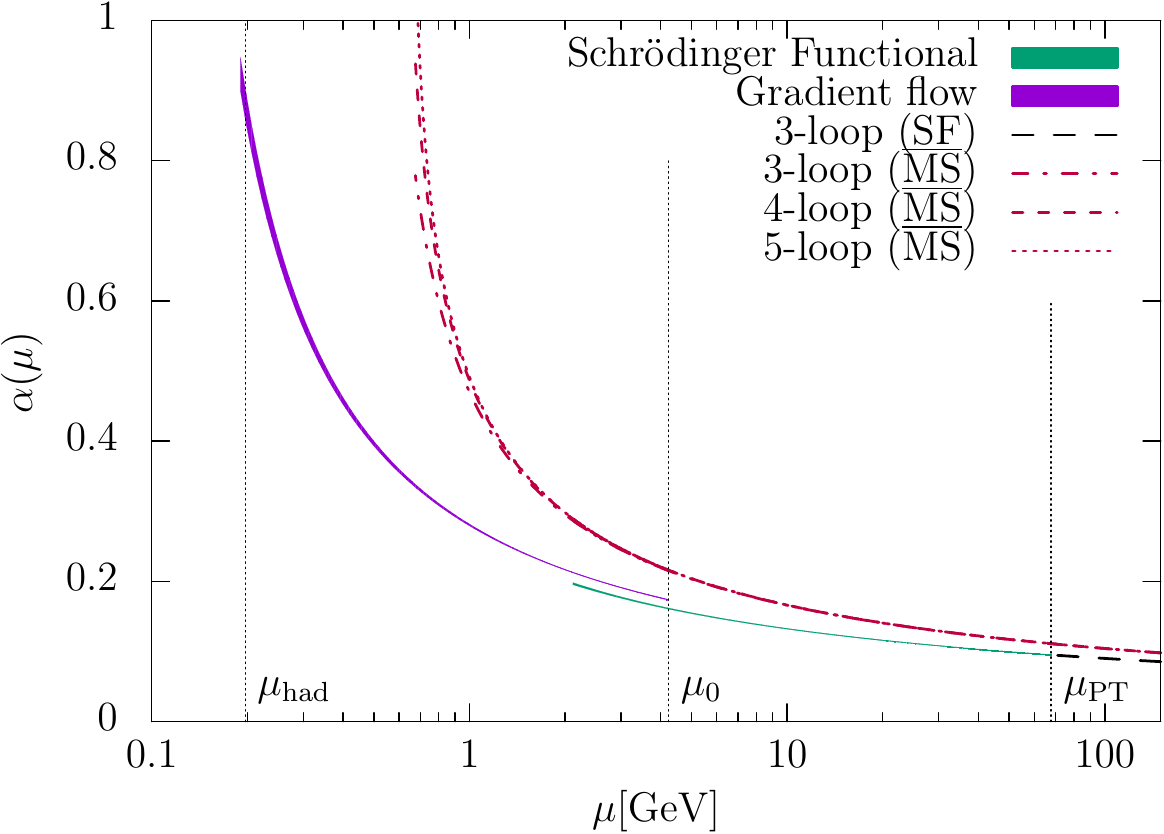}
  \caption{\label{fig:alpha}
           Running couplings of $\nf=3$ QCD from integrating the
           nonperturbative $\beta$-functions in the SF and GF
           schemes~\cite{Brida:2016flw,DallaBrida:2016kgh}. They are matched
           nonperturbatively by defining $\gbar_{\rm SF}^2(\mu_0) = 2.012$ and
           computing $\gbar_{\rm GF}^2(\mu_0/2) = 2.6723(64)$.
          }
\end{figure}

In order to exploit the advantages of both finite-volume schemes, we use  the
GF scheme at low energies, between $\muhad$ and $\mu_0$.  There we switch
nonperturbatively to the SF scheme (see Figure~\ref{fig:alpha}). Then we run up
to $\mupt$.  In this way, we connected hadronic scales to
$\mupt$~\cite{Brida:2016flw,DallaBrida:2016kgh}, cf. Table~\ref{tab:scales}.

\begin{table}
  \caption{\label{t:nprunning}
           Scale ratios and values of the coupling determined from
           nonperturbative running from $\muhad{}$ to $\mu_0/2$ in the GF and
           from $\mu_0$ to $\mupt$ in the SF scheme.
          }
  \begin{ruledtabular}
  \begin{tabular}{llll}
    $\gbar_{\rm GF}^2(\muhadi{})$& $\gbar^2_{\rm SF}(\mupt)$  & $\mupt/\muhad{}$ & $\Lambda_\msbar^{(3)}/\muhad{}$ \\
    \midrule
    11.31& 1.193(5)  & 349.7(6.8) &  1.729(57) \\
    10.20& 1.193(5)  & 322.2(6.3) &  1.593(53)
  \end{tabular}
  \end{ruledtabular}
\end{table}

In Table~\ref{t:nprunning} we show our intermediate results for $\gbar_{\rm
SF}^2(\mupt)$ and $\mupt/\muhadi{}$ for two choices\footnote{In \cite{DallaBrida:2016kgh} only $\muhadi{,1}$ was considered. Here
we extend the analysis to $\muhadi{,2}$ in order to have an additional check of
our connection of large and small volume physics.}
of a typical hadronic scale $\muhad{}$ of a few hundred {\rm MeV}.  In
addition, we give $\Lambda_\msbar^{(3)}/\muhad{}$,  obtained by the NNLO
perturbative asymptotic relation and the
exact conversion to the $\msbar$ scheme.  We have verified that the systematic
uncertainty $\sim\alpha^2(\mupt)$ and power corrections $\sim(\Lambda/\mupt)^k$
from this limited use of perturbation theory at scales above $\mupt$ are
negligible compared to our statistical uncertainties \cite{Brida:2016flw,paperSF2017}.

\section{Connection to the hadronic world}\label{sec:had}
The second key element is the nonperturbative determination of $\mu_{\rm had}$
in units of the experimentally accessible $\fpik$.  Our determination is based
on CLS ensembles \cite{Bruno:2014jqa} of $\nf=3$ QCD with
$\mup=\mdown\equiv\mlight$ in large volume.  It is convenient to define a scale
$\muref$ by the condition\footnote{Note that $\muref$ is defined ensemble by ensemble, and therefore  it is
a function of the quark masses. Instead of $\mu_{\rm ref}$, it is customary in
the lattice literature to quote $\sqrt{8t_0}=1/\mu_{\rm ref}$ 
\cite{Luscher:2010iy}.
}
\begin{equation}
    \gbar_{\infty}^2(\muref) = 1.6\pi^2 \approx 15.8\,,
\end{equation}
and trajectories in the (bare) quark mass plane $(\mlight,\mstrange)$ by
keeping the dimensionless ratio
\begin{equation}
    \phi_4 = (\mk^2 +\mpi^2/2) \, /\, \muref^2\,
    \label{e:phi4}
\end{equation}
constant.  Moreover, we define a reference scale $\mu_{\rm ref}^\star$ at the
symmetric point ($\mup=\mdown=\mstrange$) by
\bes
    \muref^\star \equiv \muref\Big\vert_{\phi_4=1.11,\, \mup=\mdown=\mstrange}\,.
  \label{e:t0star}
\ees
The requirement that the $\phi_4$=constant trajectory 
passes through the physical point, defined by
\bes
  \mpi^2/\fpik^2 = 0.8341, \quad
  \mk^2/\fpik^2  = 11.21\, , 
\ees
results in $\phi_4=1.11(2)$ in the continuum limit \cite{Bruno:2016plf} and motivates 
the particular choice in \eq{e:t0star}.

\begin{figure}
  \vspace*{-1mm}
  \includegraphics[width=0.95\columnwidth]{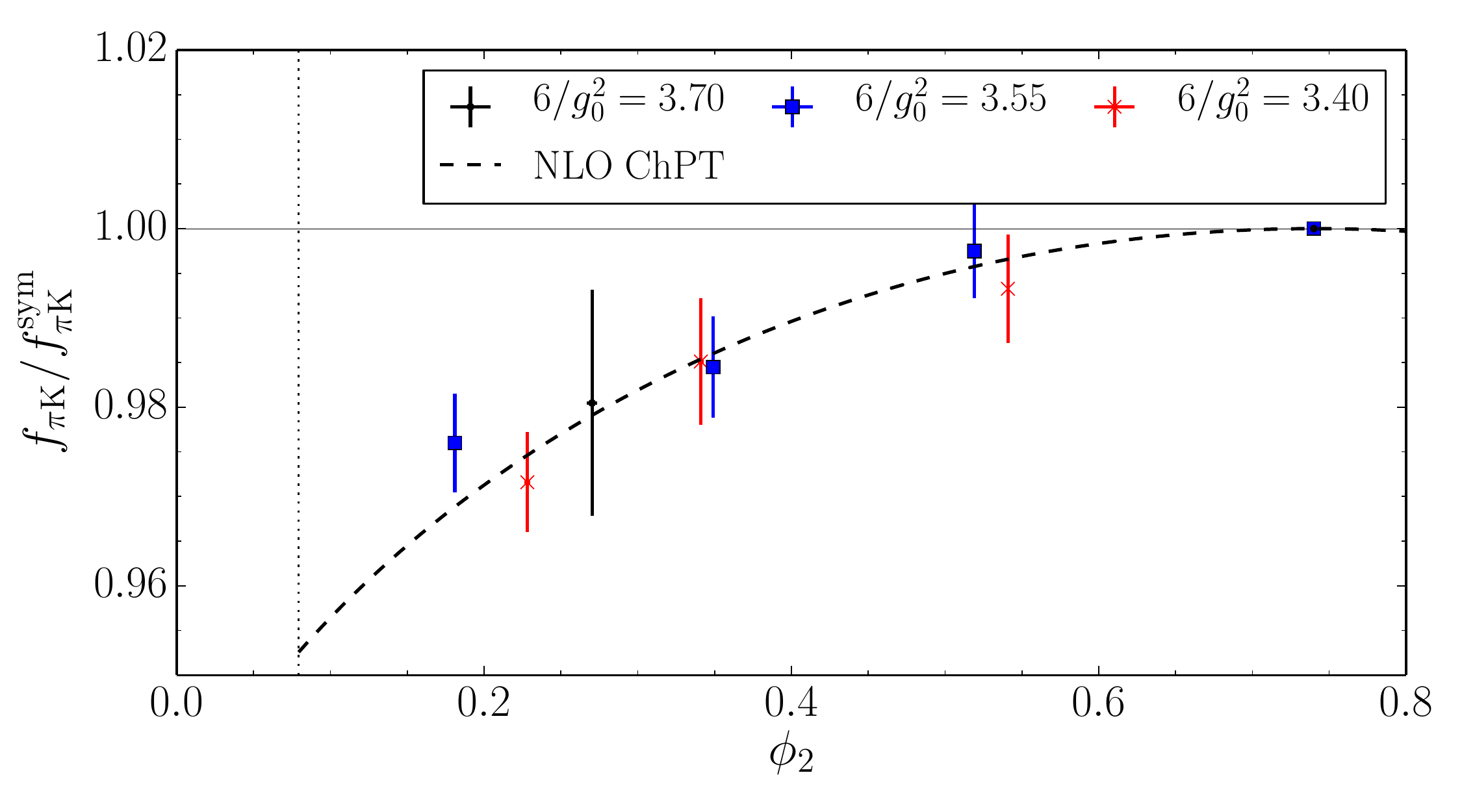}
  \caption{\label{f:chi}
           Dependence of $\fpik$ at $\phi_4=1.11$ on the pion mass through
           $\phi_2=m_\pi^2/\muref^2$~\cite{Bruno:2016plf}.  We normalized to
           $\fpik^{\rm sym}$ at the symmetric point $\mup=\mdown=\mstrange$.
           The ratio follows the parameter-free prediction of NLO chiral PT.
              }
\end{figure}

Since the combination $\fpik$ has a weak and well understood dependence on the
pion mass along trajectories with constant $\phi_4$, a precise extrapolation
from the symmetric point to the physical point can be performed
\cite{Bietenholz:2010jr,Bruno:2016plf}, see Figure~\ref{f:chi}.  Continuum
extrapolations with four lattice spacings, $0.05\, \fm  \lesssim a \lesssim
0.09\, \fm$, together with the PDG value of \eq{e:fpikpdg},
yield~\cite{Bruno:2016plf}\bes
  \muref^\star = 478(7) \, {\rm MeV} \,.
  \label{e:t0starfm}
\ees
Note that $\mu_{\rm ref}^\star$ is defined at a point with {\em unphysical}
quark masses, where finite-size effects are smaller and simulations are easier
than close to the physical point. This allows us to include in the following
analysis a CLS ensemble at a fifth lattice spacing, $a \approx 0.039\,\fm$.

For the determination of $\muref^\star/\muhad$, we need pairs of values
$a\muhad{}$ and $a\muref^\star$ at the same value of $a$.  This requires either
an interpolation of the data for $a\mu_{\rm had}$, or an interpolation of the
data for $a\mu_{\rm ref}^\star$. We denote these two options as set \texttt{A}
and \texttt{B}, respectively. 

\begin{figure}[htb!]
   \includegraphics[width=\columnwidth]{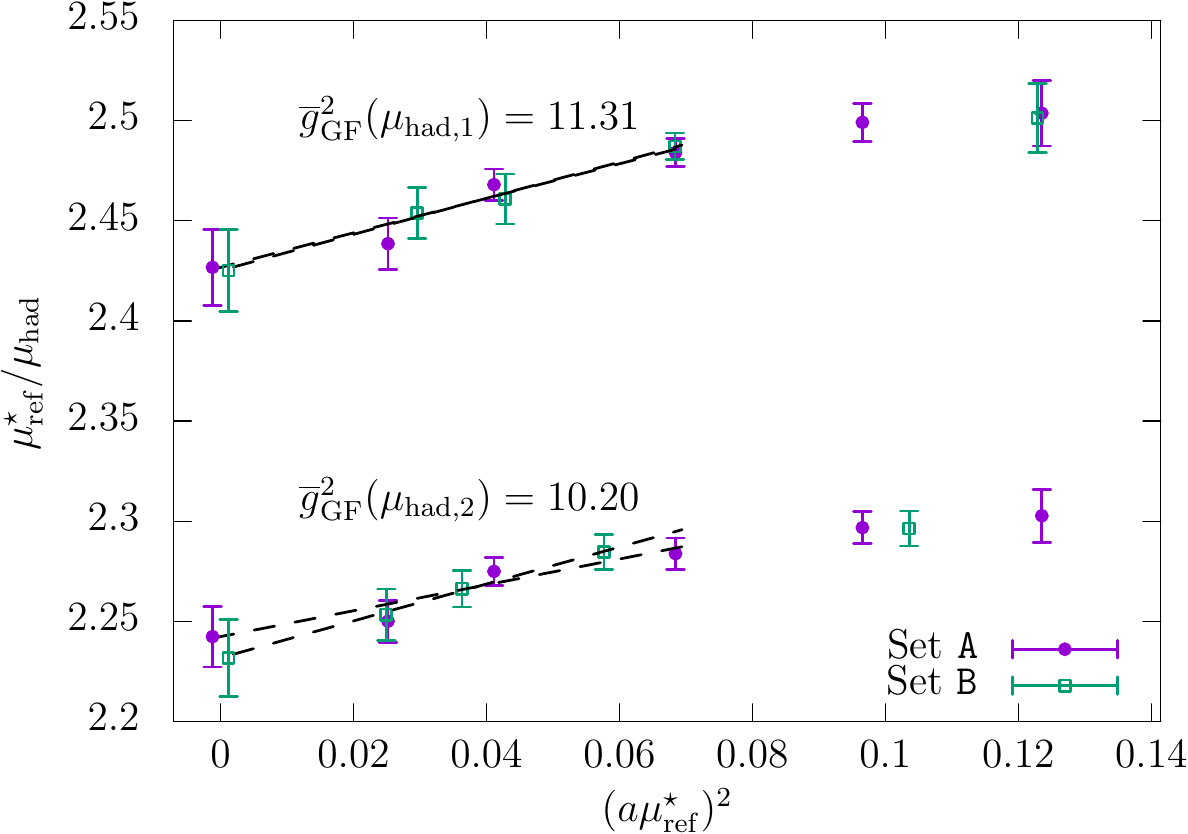}
   \caption{\label{f:contextrap1}
            Continuum extrapolations described in the text. Extrapolated
            values are shown in proximity of $a=0$.
           }
\end{figure}

The dimensionless ratio $\muref^\star/\muhadi{}$  can then be extrapolated to
the continuum as shown in Figure~\ref{f:contextrap1}.  Extrapolations, linear
in $a^2$  dropping data above $(a\muref^\star)^2 =0.07$  with either set
\texttt{A} or \texttt{B}, are fully compatible. They are also stable under
changes in the number of points used to extrapolate and the particular
functional form. These stabilities are expected since our smallest lattice
spacing is $a\approx 0.039\,\fm$. 
We repeat the computation of $\muref / \mu_{\rm had}$ for two different values of
$\mu_{\rm had}$ (see Figure 4).
Tables of the various numbers that enter and further
details can be found in the supplementary material.

As our final estimates we take set \texttt{B}, which has somewhat larger
errors, and obtain
\bes
\frac{\muref^\star}{\muhadi{,1}} = 2.428(18)\,, \quad 
\frac{\Lambda_\msbar^{(3)} }{\muref^\star} = 0.712(24) \,,
  \nonumber \\[-1ex]
  \\[-1ex]
  \nonumber
\frac{\muref^\star}{\muhadi{,2}} = 2.233(17)\,, \quad 
\frac{\Lambda_\msbar^{(3)} }{\muref^\star} = 0.713(24) \,.
\ees
The close agreement in $\Lambda_\msbar^{(3)} / \muref^\star$ is reassuring.
With \eq{e:t0starfm} we arrive at our central result
\bes
    \label{e:lamda3res}
   \Lambda_\msbar^{(3)} = 341 (12)\,\MeV\,.
\ees
It has a remarkable precision given that we ran the couplings nonperturbatively
up to about $70\,\GeV$ and only then used perturbation theory.

\section{$\Lambda$ parameters and coupling of five- and six-flavor theories}
By itself our $\Lambda^{(3)}$  is of limited phenomenological use. The
three-flavor effective field theory (EFT) is valid for energies below
$m_\mathrm{charm}=1.28\, {\rm GeV}$.  There perturbation theory cannot be
expected to be precise.

However, QCD$^{(\nf)}$, the $\nf$-flavor effective theory,  can be matched to
QCD$^{(\nf+1)}$ and one can eventually arrive at QCD$^{(6)}$
\cite{Weinberg:1951ss}. This matching relates the couplings such that the low(er) energy EFT agrees with the
(more) fundamental one up to power law corrections. These
$\rmO(\Lambda^2/m_\mathrm{h}^2)$ corrections can only be studied
nonperturbatively. They are very small already for $m_\mathrm{h}=\mcharm$
\cite{Bruno:2014ufa,Korzec:2016eko}.

Ignoring $1/m_{\mathrm{h}}^2$ effects, matching means
\bes
    \label{e:matchgbar}
     \gbarnf(\mu)=\gbarpl(\mu) \, \times \,
         \xi\left(\gbarpl(\mu), \frac{m_\mathrm{h}}{\mu}\right)
         \nonumber  \\[-1ex]
\ees         
and the $\Lambda$ parameters are related by
\bes
    \label{e:lamrat}
    \frac{\Lam^{(\nf)}}{\Lampl}  
    =  
    \frac{\varphi^{(\nf)}(\gbar^{(\nf+1)} \times \xi)}{\varphi^{(\nf+1)}(\gbar^{(\nf+1)})} \,,
\ees 
where 
\begin{eqnarray}
    \varphi^{(\nf)}(\gbar) &=& ( b_0 \gbar^2 )^{-b_1/(2b_0^2)} 
        \rme^{-1/(2b_0 \gbar^2)} \label{e:phig}  \\
     && \times \exp\left\{-\int\limits_0^{\gbar} \rmd x\ 
        \left[\frac{1}{\beta(x)} 
             +\frac{1}{b_0x^3} - \frac{b_1}{b_0^2x} \right] \right\} \, 
        \nonumber  
\end{eqnarray}
is defined in terms of the $\nf$-flavor $\beta$-function in the chosen scheme.

\begin{table}
  \caption{\label{tab:res}
           Main results of this work. $\Lambda_\msbar^{(3)}$, \eq{e:lamda3res}, 
           is determined
           nonperturbatively and used, together with the perturbative estimates
           of $\Lambda_\msbar^{(\nf)}/\Lambda_\msbar^{(3)}$, to produce all
           other numbers. 
           As additional input we use the masses 
            $\mcharm^*       =  1.280(25)  \,\GeV$,                  $\mbeauty^*      =  4.180(30)  \,\GeV$,                              $m_\mathrm{Z}    = 91.1876\,\GeV$,    \cite{Olive:2016xmw} and
            $\mtop^*         = 165.9(2.2)\,\GeV$, \cite{Fuster:2017rev}.
            } 
  \begin{ruledtabular}
  \begin{tabular}{llll} 
    $\nf$ & $\Lambda^{(\nf)}_\msbar$ [\MeV] & $\mu$ & $\alpha_\msbar^{(\nf)}(\mu)$ \\
    \midrule
    4 & 298(12)(3)& \\     5 & 215(10)(3)        & $m_\mathrm{Z}$     & 0.11852(80)(25)    \\
    6 & 91.1(4.5)(1.3) & $\mulhc$              & 0.08523(41)(12) \\
  \end{tabular}
  \end{ruledtabular}
\end{table}

When inserting the perturbative expansions of $\xi$ and $\beta$, we choose the
mass $m_\mathrm{h}$ in \eq{e:matchgbar} as the $\msbar$ mass at its own scale,
$m^*=\mbar_\msbar(m^*)$, and set $\mu=m^*$.  Then the one-loop term vanishes in
the perturbative expansion
\bes        
   \xi(\gbar,1) = 1+c_2\,\gbar^4+c_3\,\gbar^6 + c_4\,\gbar^8 + \rmO(\gbar^{10})\,. 
   \label{e:match} 
\ees
For numerical results, we use $c_2,c_3,c_4$
\cite{Weinberg:1980wa,Bernreuther:1981sg,Grozin:2011nk,Chetyrkin:2005ia,Schroder:2005hy} together 
with the appropriate five-loop $\beta$-function~\cite{MS:4loop1,Czakon:2004bu,Baikov:2016tgj,Luthe:2016ima,Herzog:2017ohr} to arrive
at Table~\ref{tab:res}.  

The first error in $\alpha(\mu)$ is due to $\Lambda_\msbar^{(3)}$ and the quark mass
uncertainties, where the latter are hardly noticeable. 
The second error listed
represents  our estimate of the truncation error in PT in the connection
$\Lambda_\msbar^{(3)}\to\Lambda_\msbar^{(4)}\,-\,\Lambda_\msbar^{(6)}$. We arrive
at it as follows.  The $2,3,4$-loop terms in \eq{e:match} combined with the
$3,4,5$-loop $\beta$-functions in \eqref{e:phig} lead, e.g.,  to contributions
$128,\,19,\,6$ in units of $10^{-5}$ to  $\alpha_\msbar(m_{\rm Z})$.  We take
the sum of the last two contributions as our perturbative uncertainty. 
Within PT, this is conservative. 
Recently, Herren and Steinhauser~\cite{Herren:2017osy}  considered also $\mu\ne
m$ in \eq{e:matchgbar}.  Their error estimate, $0.0004$, would 
change little in the uncertainty of our final result 
\bes
  \alpha_\msbar^{(5)}(m_\mathrm{Z}) = 0.11852(84) \,.
\ees

\section{Summary and Conclusions}

QCD offers a plethora of quantities, like hadron masses and meson decay
constants, that can be used as precise experimental input to compute the strong
coupling and quark masses. However, the nonperturbative character of the strong
interactions makes these computations difficult. Lattice QCD offers a
unique tool to connect, from first principles, well-measured QCD
quantities at low energies to the fundamental parameters of the Standard Model. As perturbative expansions are not convergent, but only asymptotic, the
challenge for precise results is to nonperturbatively reach energy scales where
the strong coupling is small enough~\cite{Brida:2016flw}. Due to the slow
running of $\alpha_s$, the hadronic and perturbative regimes are separated by two to three orders of magnitude. 

Finite-size scaling allows one to bridge such large energy differences
nonperturbatively. 
It trades the systematic uncertainties associated with
the truncation of the perturbative series at relatively low energies
for statistical uncertainties which are easy to estimate.

Our precise data for the running
coupling~\cite{Brida:2016flw,DallaBrida:2016kgh}, together with the
high-quality set of ensembles provided by the CLS
initiative~\cite{Bruno:2014jqa} at lattice spacings as small as $a\approx
0.039\,\fm$, and an accurate determination of the scale~\cite{Bruno:2016plf},
allow us to reach a precision of 0.7\% in $\alpha_\msbar^{(5)}(m_\mathrm{Z})$.

The factor $\mu_{\rm PT}/\mu_{\rm had}$ contributes 87\% of the uncertainty in
$\alpha_\msbar^{(5)}$. This uncertainty is dominantly statistical and could 
certainly be reduced significantly by some additional effort. While 
present knowledge
indicates small and perturbatively computable quark-loop effects in the
matching at the heavy-quark thresholds, the uncomfortable need of using PT
at scales as low as $m_\mathrm{charm}$ can only be avoided by a full
four-flavor computation. This is a mandatory step as soon as one
attempts another controlled reduction of the total uncertainty.

We finally note, that our result $\alpha_\msbar^{(6)}(\mulhc)= 0.0852(4)$ is
in good agreement with the recent CMS determination \cite{Khachatryan:2016mlc} from jet cross sections with $p_\mathrm{T} \in [1.41,2.5]\,\TeV$.
Ref.~\cite{Khachatryan:2016mlc} gives
$\alpha_\msbar^{(5)}(1.508\,\TeV)=0.0822(33)$ 
which was already converted to
$\alpha_\msbar^{(6)}(1.508\,\TeV)=0.0840(35)$ in
\cite{Herren:2017osy}.
Although LHC data does not yet reach the
precision of our result (evolved from lower energy),
comparisons at such high energies are an excellent test of QCD and of the
existence of massive colored quanta.

\vskip1em
\begin{acknowledgments}
\emph{Acknowledgments---}The technical developments which enabled the results presented in this paper
are based on seminal ideas and ground breaking work by Martin L\"uscher, Peter
Weisz and Ulli Wolff; most importantly, the use of finite-size scaling methods
for renormalized couplings, perturbation
theory on the lattice and in the SF to two loop
order, and the use of the gradient
flow. We would like to express our gratitude to Martin,
Peter and Ulli for collaborative work, numerous enlightening discussions and
advice over the years. Furthermore, we thank our colleagues in the ALPHA
collaboration for helpful feedback, P.~Marquard and P.~Uwer for discussions
on the $\msbar$ top quark mass,
and M. Steinhauser and F. Herren for correspondence on 
refs.~\cite{Herren:2017osy,Khachatryan:2016mlc}.

We thank our colleagues in the Coordinated Lattice Simulations (CLS) effort 
[\url{http://wiki-zeuthen.desy.de/CLS/CLS}] for the joint generation of the gauge field ensembles on which the computation described here is based.
 We acknowledge PRACE  for awarding us access to resource 
FERMI (projects ``LATTQCDNf3'', Id 2013081452 and
``CONTQCDNf3'', Id 2015122835), 
based in Italy, at CINECA, Bologna 
and to resource SuperMUC  (project ``ContQCD", Id 2013081591),
based in Germany at LRZ, Munich.
We  are grateful for the support received by the computer centers.

The authors gratefully acknowledge the Gauss Centre for Supercomputing (GCS)
for providing computing time through the John von Neumann Institute for
Computing (NIC) on the GCS share of the supercomputer JUQUEEN at J\"ulich
Supercomputing Centre (JSC). GCS is the alliance of the three national
supercomputing centers HLRS (Universit\"at Stuttgart), JSC (Forschungszentrum
J\"ulich), and LRZ (Bayerische Akademie der Wissenschaften), funded by the
German Federal Ministry of Education and Research (BMBF) and the German State
Ministries for Research of Baden-W\"urttemberg (MWK), Bayern (StMWFK) and
Nordrhein-Westfalen (MIWF).

We thank the computer centers at HLRN (bep00040), NIC at DESY Zeuthen and CESGA
at CSIC (Spain) for providing computing resources and support. 
M.B. was supported by the  U.S. D.O.E. under Grant No. DE-SC0012704. 
M.D.B. is grateful to CERN for the hospitality and support.
S.Si.  acknowledges support by SFI under grant 11/RFP/PHY3218.
This work is based on previous work \cite{Sommer:2015kza} supported strongly by
the Deutsche Forschungsgemeinschaft in the SFB/TR~09. 

\end{acknowledgments}

\section{App: Supplementary material}

Here we give some additional details concerning the new computation described
in the section {\it Connection to the hadronic world}. 

The continuum extrapolation of $\muref^\star/\muhad$ requires values of
$a\muhad$ and $a\muref^\star$ at identical values of the improved bare
coupling $\tilde g_0$ (as defined below).
Thus, the determination of $\muref^\star/\muhad$ consists of three main parts
\begin{itemize}
  \item[(I)]   $a\muhad$ vs. $g_0$ from simulations in finite volume
  \item[(II)]  $a\muref^\star$ vs. $\tilde g_0$ from simulations in large volume
  \item[(III)] determination of $\muref^\star/\muhad$ 
\end{itemize}
which we now describe in detail.

\section{(I) $a\muhad$ vs. $g_0$ in finite volume}

\subsection{Lines of constant physics}

A smooth continuum limit requires precise definitions of the renormalized
parameters that are kept fixed while sending $a \to 0$. This defines the
``lines of constant physics'' for the choice of the bare parameters in the
simulations.
Since $\muhad$ is defined for massless quarks and in a finite volume with
$\mu=1/L$, we choose the conditions
\begin{equation}\label{e:LCP1}
    m(\muhadi{,\it i}) = 0 \,,
\end{equation}
where $m(\mu)$ is the current quark mass defined through suitable Schr\"odinger
Functional correlators~\cite{Luscher:1996sc}, and
\begin{equation}\label{e:LCP2}
   \bar{g}^2_{\rm GF}(\muhadi{,\it i}) = u_{{\rm had,}i} \,,
\end{equation}
with our values $u_{{\rm had,}1}=11.31$ and $u_{{\rm had,}2}=10.20$ for $\muhadi{,1}$ and
$\muhadi{,2}$, respectively.

For a given $g_0$ and lattice size $L/a$, \eq{e:LCP1} is equivalent to the
determination of the \emph{critical} values of the bare quark masses, 
$am_{\rm cr}(g_0,a/L)$, at which the renormalized quark masses vanish. For
later convenience it is useful to parametrize the deviation from the critical
line by introducing the subtracted quark mass: $am_{\rm q}= am_0-am_{\rm
cr}(g_0,a/L)$, where 
$m_{{\rm u},0}=m_{\rm d,0}=m_{{\rm s},0}=m_0$~\cite{Luscher:1996sc} are
the bare mass parameters of the lattice action.

The exact definition of $m$, i.e., the kinematical choices in the
implementation of the  PCAC relation,  is given in \cite{DallaBrida:2016kgh}. This
leads to the results for $am_{\rm cr}(g_0,a/L)$ reported in section~A.1.4,
specifically eq.~(A.3) of this reference. This determination is
available for $L/a=8,12,16$, and for the whole range of $g_0$ values we considered.
In particular, these values of $m_{\rm cr}$ guarantee that $|Lm(1/L)|<0.005$,
which is a stringent enough bound to let us assume that $m(1/L)$ is effectively
zero in our analysis. We note, however, that below we shall consider also
lattices with $L/a=20,24,32$ (and larger). For these, $am_{\rm cr}(g_0,a/L)$ at
a given $g_0$ was estimated by performing a linear extrapolation in $(a/L)^3$
of the critical hopping parameter, $\kappa_{\rm cr}(g_0,a/L)=(2am_{\rm
cr}(g_0,a/L)+8)^{-1}$, using the results at the two largest available lattices,
namely $L/a=12,16$.  Examples of these kind of extrapolations are explicitly
discussed in~\cite{Nf3_tuning}.

\subsection{Simulations and results for $a\muhad$}

\begin{table*}[hbpt]
  \small\footnotesize
  \caption{\label{tab:SetEnsembles}
           List of all finite-volume ensembles that enter the hadronic matching
           computation.  Here $\kappa=(2am_0+8)^{-1}$.  The column labeled by
           $\bar{g}^2_{\rm GF}$ contains the values of the GF coupling as
           measured on the given ensembles. The values for the mass derivative
           ${\rm d} \bar{g}^2_{\rm GF}/{\rm d}am_{\rm q}$ instead are only
           given for ensembles with $am_{\rm q}=\Delta am_{\rm cr}\neq0$. The
           coupling resulting from the shift to $am_{\rm q}=0$ is listed in
           the last column.
          }
  \begin{ruledtabular}
  \begin{tabular}{llldccd}
  $L/a$ & $6/g_0^2$  & $\kappa$            &  \multicolumn{1}{c}{$\bar{g}^2_{\rm GF}$}
                                                        & ${\rm d}\bar{g}^2_{\rm GF}/{\rm d}am_{\rm q}$ 
                                                                      & $am_{\rm q}$          &  \multicolumn{1}{r}{$\bar{g}^2_{{\rm GF}, am_{\rm q}=0}$}  \\
   \midrule
   $12$ & $3.400000$ & $0.136872258739141$ & 11.308(97) & $-$         & $0$                   & 11.308(97) \\
   $12$ & $3.480000$ & $0.137038980000000$ &  9.417(41) & $-$         & $0$                   &  9.417(41) \\
   $12$ & $3.497000$ & $0.137062990000000$ &  9.118(54) & $-$         & $0$                   &  9.118(54) \\
   $12$ & $3.500000$ & $0.137066849094359$ &  9.104(28) & $-$         & $0$                   &  9.104(28) \\
   $12$ & $3.510000$ & $0.137078900000000$ &  8.897(46) & $-$         & $0$                   &  8.897(46) \\
   $12$ & $3.532000$ & $0.137101170000000$ &  8.738(40) & $-$         & $0$                   &  8.738(40) \\
   $12$ & $3.547000$ & $0.137113150000000$ &  8.460(38) & $-$         & $0$                   &  8.460(38) \\
   \midrule
   $16$ & $3.530000$ & $0.137142109937020$ & 11.917(89) & $-$         & $0$                   & 11.917(89) \\
   $16$ & $3.540000$ & $0.137148520500305$ & 11.565(82) & $-$         & $0$                   & 11.565(82) \\
   $16$ & $3.556470$ & $0.137032452074134$ & 11.54(12)  & $ 87(10)  $ & $3.31 \times 10^{-3}$ & 11.25(11)  \\
   $16$ & $3.556470$ & $0.137032452074134$ & 11.39(11)  & $ 87(12)  $ & $3.31 \times 10^{-3}$ & 11.105(99) \\
   $16$ & $3.560696$ & $0.137036459478077$ & 11.267(76) & $ 83.2(76)$ & $3.25 \times 10^{-3}$ & 10.996(71) \\
   $16$ & $3.560696$ & $0.137158587623782$ & 10.982(64) & $-$         & $0$                   & 10.982(64) \\
   $16$ & $3.629800$ & $0.137163450000000$ &  9.638(34) & $-$         & $0$                   &  9.638(34) \\
   $16$ & $3.653850$ & $0.137072212042003$ &  9.250(69) & $ 34.9(62)$ & $2.22 \times 10^{-3}$ &  9.173(66) \\
   $16$ & $3.657600$ & $0.137154130000000$ &  9.169(49) & $-$         & $0$                   &  9.169(49) \\
   \midrule
   $20$ & $3.682900$ & $0.137147800000000$ & 11.404(73) & $-$         & $0$                   & 11.404(73) \\
   $20$ & $3.790000$ & $0.137048000000000$ &  9.251(54) & $-$         & $0$                   &  9.251(54) \\
   \midrule
   $24$ & $3.735394$ & $0.137082625274551$ & 12.87(16)  & $126(18)  $ & $6.51 \times 10^{-4}$ & 12.79(16)  \\
   $24$ & $3.793389$ & $0.137020768592807$ & 11.79(12)  & $101(13)  $ & $5.16 \times 10^{-4}$ & 11.74(12)  \\
   $24$ & $3.833254$ & $0.136967740552669$ & 10.497(78) & $ 32.7(62)$ & $4.47 \times 10^{-4}$ & 10.482(78) \\
   $24$ & $3.936816$ & $0.136798051283124$ &  8.686(52) & $ 19.2(40)$ & $3.30 \times 10^{-4}$ &  8.680(52) \\
   \midrule
   $32$ & $3.900000$ & $0.136872019362733$ & 13.36(15)  & $111(28)$   & $1.53 \times 10^{-4}$ & 13.34(15)  \\
   $32$ & $3.976400$ & $0.136730873919691$ & 11.34(11)  & $ 61(22)$   & $1.27 \times 10^{-4}$ & 11.34(11)  \\
   $32$ & $4.000000$ & $0.136683960224116$ & 10.91(13)  & $ 56(18)$   & $1.22 \times 10^{-4}$ & 10.91(12)  \\
   $32$ & $4.100000$ & $0.136473008507319$ &  9.077(80) & $ 29(14)$   & $7.39 \times 10^{-5}$ &  9.075(78) \\
  \end{tabular}
  \end{ruledtabular}
\end{table*}

\begin{figure}[b]
  \includegraphics[width=\columnwidth]{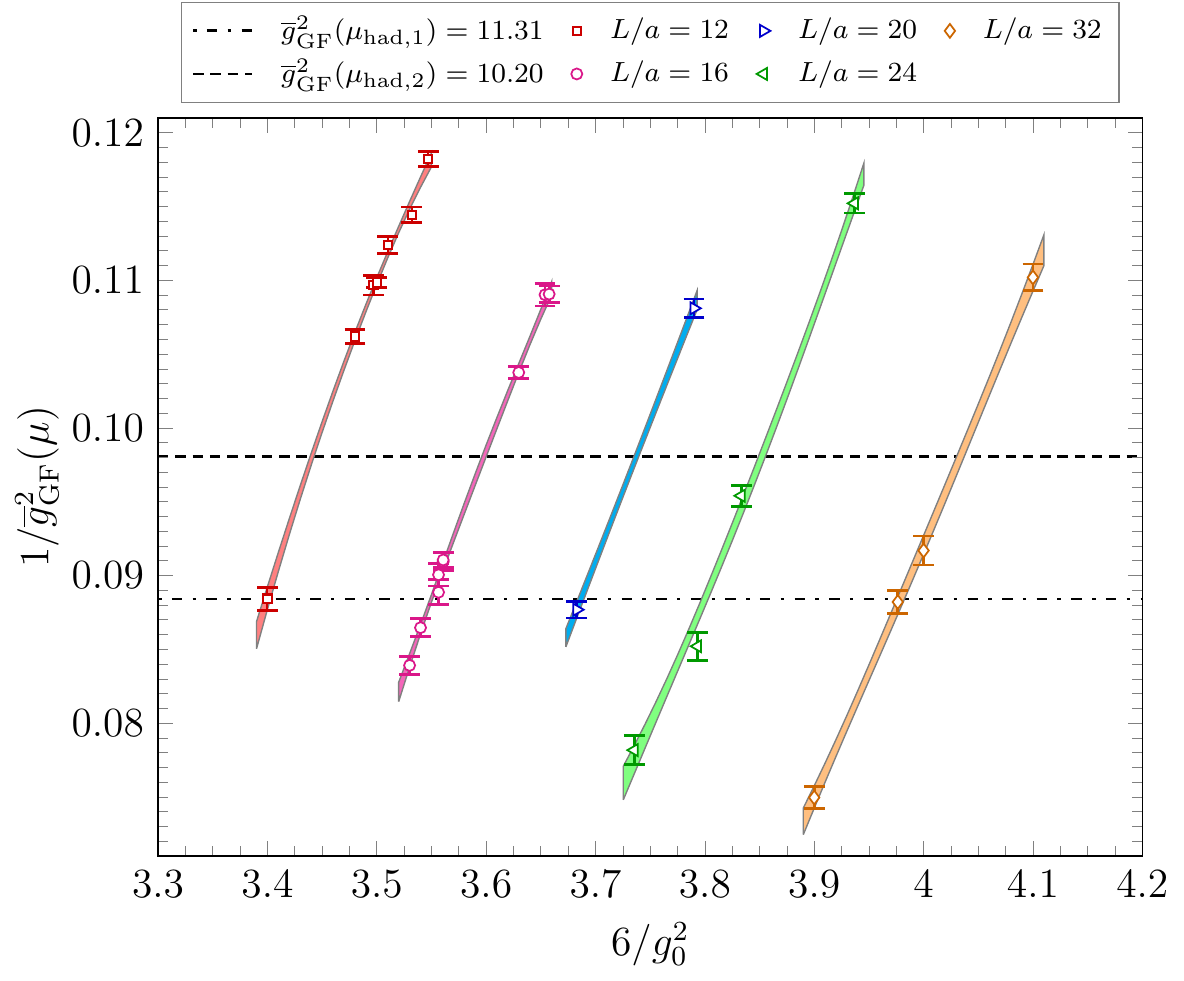}
  \caption{\label{f:uinvfit}
           Interpolations $\gbar_{{\rm  GF}}^{-2} = k_0 + k_1 (6/g_0^2) +k_2
           (6/g_0^2)^2$, performed separately for each $L/a$, with $k_2=0$ for
           $L/a=20$.
          }
\end{figure}

To perform the hadronic matching we collected many different ensembles with
$L/a\in\{12,16,20,24,32\}$, at several values of $g_0$. These were used to
determine pairs $(g_0,L/a)$ for which the conditions (\ref{e:LCP1}) and
(\ref{e:LCP2})  are satisfied. This is discussed in detail in the next section.
The full set of ensembles is given in Table \ref{tab:SetEnsembles}. For more
than half of the ensembles the bare quark masses are such that $am_{\rm q}=0$,
according to the definition of $am_{\rm cr}(g_0,a/L)$ described above. For the
others instead, we have $am_{\rm q}=\Delta am_{\rm cr}= am_{\rm cr}(g_0,2a/L)-am_{\rm
cr}(g_0,a/L)$, because these ensembles originated as \emph{doubled} lattices
in the step scaling study of ref.~\cite{DallaBrida:2016kgh}.  However, in order
to have a smooth $\Or(a^2)$ dependence of the cutoff effects of our observables
along lines of constant physics, all ensembles should have $m_{\rm q}=0$
according to a unique and specific definition (see e.g.
ref.~\cite{Luscher:1996jn}). 

In order to achieve this, we computed in our MC-simulations the derivatives:
\begin{equation}
 \frac{{\rm d} \bar{g}^2_{\rm GF}(\mu)}{{\rm d}am_{\rm q}}
\end{equation}
on the ensembles with $am_{\rm q}\ne 0$, cf. Table~\ref{tab:SetEnsembles}. This information was
then used to correct the measured values of $\bar{g}^2_{\rm GF}(\mu)$ on these
ensembles in order to fulfill the condition $m_{\rm q}=0$. 
With the resulting values, $\gbar_{{\rm GF},am_{\rm q}=0}^2$, we can safely 
perform a smooth
interpolation of the data for $\gbar_{{\rm GF},am_{\rm q}=0}^2$ to the target
values $u_{{\rm had,}1}=11.31$ and $u_{{\rm had,}2}=10.20$. 

At fixed $L/a$ we fit the data for  $ \gbar_{{\rm  GF},am_{\rm q}=0}^{-2}$ as a
function of $6/g_0^2$ using a second degree polynomial, except for the data at
$L/a=20$ where a linear interpolation is used. These fits are all of excellent
quality, except for $L/a=24$,  see Figure~\ref{f:uinvfit}.\footnote{We have of course checked the $L/a=24$ simulations very carefully for 
          autocorrelation and thermalization effects and have come to the
          conclusion that the poor fit quality is a case of a large statistical
          fluctuation.  In all cases (including for $L/a=24$), our interpolated
          pairs $(g_0, a\mu_{\rm had})$ are stable under a change in the
          functional form used to fit the data, e.g., a Pad\'e ansatz or a
          global fit where the coefficients of the polynomial are parametrized
          as smooth functions in $a/L$.
} 
Our final interpolated pairs for $(g_0, a\mu_{\rm had})$ are shown in
Table~\ref{t:t0lmax1}, set \texttt{B}. 

The improved bare coupling ${\tilde g_0}$ is defined
as~\cite{Luscher:1996sc,Bhattacharya:2005rb}
\begin{equation}
{\tilde g_0^2} = g_0^2\left(1+\tfrac{1}{3}{\rm tr}\{aM_{\rm q}\}\, b_{\rm
    g}(g_0)\right) \,, 
\label{e:g0tilde}
\end{equation}
where $M_{\rm q}={\rm diag}(m_{\rm u,q},m_{\rm d,q},m_{\rm s,q})$ is the
subtracted quark mass matrix and $b_{\rm g}(g_0)$ is a given function of the
bare coupling.  

As discussed above, we have $m_{\rm u,q}=m_{\rm d,q}=m_{\rm s,q}=0$
in the ensembles used for the determination of $a\mu_{\rm had}$, and 
therefore ${\tilde g_0}=g_0$.

\section{(II) $a\muref^\star$ vs. $\tilde g_0$ in large volume}

\subsection{Improved coupling in CLS simulations}

On the other hand, on the large volume CLS ensembles where $a\mu_{\rm ref}^\star$ has been
computed, the value of ${\rm tr}\{aM_{\rm q}\}$ in (\ref{e:g0tilde}) can be
estimated using the results for $am_{i,0}$ given in ref.~\cite{Bruno:2016plf},
and by taking for $am_{\rm cr}(g_0,0)$ the corresponding result from the
extrapolation of $\kappa_{\rm cr}(g_0,a/L)$ described in the previous
subsection.

Finally, at present, for the set-up of the CLS simulations, $b_{\rm g}(g_0)$ is
only known to one-loop order in perturbation theory. We thus used this value,
which is given by~\cite{Sint:1995ch}, 
\begin{equation}
  \label{e:bg}
  b_{\rm g}(g_0) = 0.036\,g_0^2 + {\rm O}(g_0^4)\,,\mbox{ for } N_{\rm f}=3\,.
\end{equation}
Due to the smallness of ${\rm tr}\{aM_{\rm q}\}$, the difference between the
bare couplings where the CLS simulations have been performed and the improved
ones (see Table~\ref{t:t0lmax1}), is very small. The ${\rm O}(g_0^4)$
uncertainty in \eq{e:bg} is irrelevant at the level of our statistical uncertainties.

Our final values for $(\tilde g_0, a\mu_{\rm ref}^\star)$ after the
correction $g_0\rightarrow \tilde g_0$ are shown in
Table~\ref{t:t0lmax1}, set \texttt{A}. There, the first four results
for $a \muref^\star$ are from \cite{Bruno:2016plf},
where also the exact definition of $\mu_{\rm ref}^\star$ at
finite $a$ is specified. The fifth number 
is from a new simulation. Some details of this interesting simulation, close
to the continuum, are given below. 

\subsection{Simulation at the symmetric point and $6/g_0^2=3.85$}

The CLS simulations, action and algorithm are described in
\cite{Luscher:2012av,Bulava:2013cta,Bruno:2014jqa,Bruno:2016plf} and the
documentation of the {\tt openQCD} code \cite{openQCD}.  To the ensembles already used in
these publications, CLS has added  a flavor-symmetric one at an even finer
lattice spacing. It is a  $192\times64^3$ lattice with open boundary
conditions, a lattice spacing of $a\approx 0.039$\,fm and a hopping parameter
of $\kappa=0.136852$.  The general algorithmic setup follows the lines of the
ones described in \cite{Bruno:2014jqa}, with only slight adjustments of the
algorithmic mass parameters due to the finer lattice spacing. 
We have used a statistics of 6k molecular dynamics units (MDU) out of the 
available 6.5k.

\begin{figure}
  \includegraphics[width=\columnwidth]{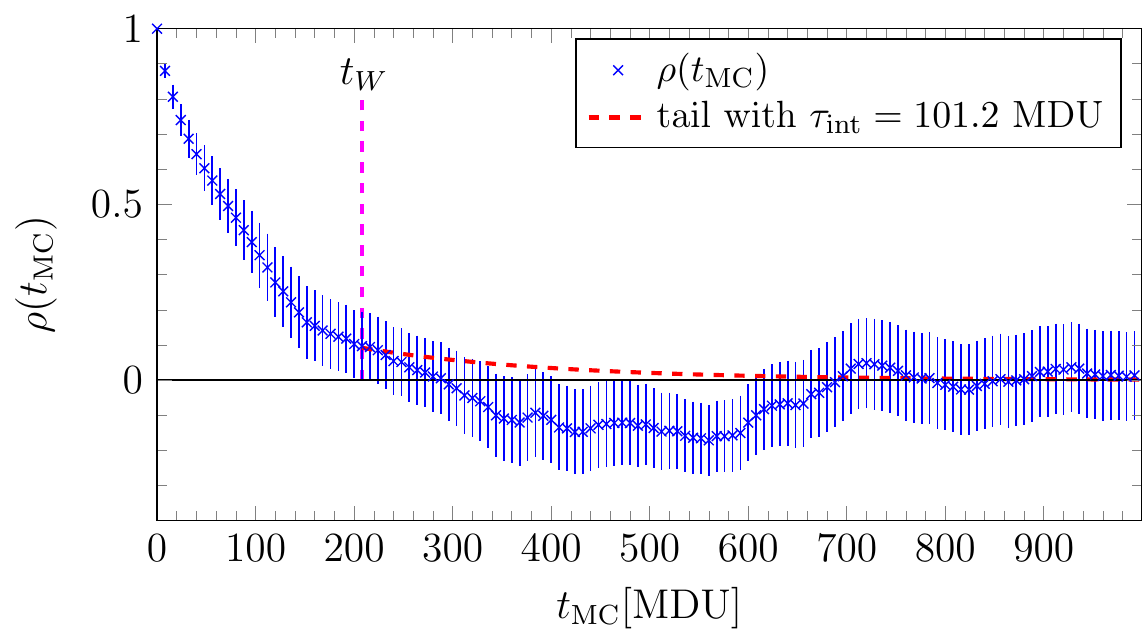}
  \caption{\label{f:rhoJ500}
           Normalized autocorrelation function, $\rho(t_\mathrm{MC})$ of
           $a\muref^\star$, as defined in~\cite{Wolff:2003sm}, and error
           estimate including a tail contribution (dashed curve), starting 
           from $t_\mathrm{W}$.
          }
\end{figure}

The estimated integrated autocorrelation time of the GF topological charge at
flow time $t=t_0$ is 0.3(2)k~MDU and  $\approx 0.1$k MDU for the action density $E(t_0)$,
cf.~(10) of the main text.  The latter is indistinguishable from $a\muref^\star$
concerning autocorrelations.  We show the normalized autocorrelation function
$\rho(t_\mathrm{MC})$ of $a\muref^\star$ in Figure~\ref{f:rhoJ500}. From
$t_\mathrm{MC}\approx$ 0.2k~MDU on, this function is compatible with zero
within relatively large errors.

For the determination of the integrated autocorrelation times, we use a
slightly modified version of the procedure proposed in  \cite{Schaefer:2010hu}.
As usual, we sum $\rho(t_\mathrm{MC})$ within a window up to $t_\mathrm{W}$, for which we take
the last $t_\mathrm{MD}$ at which $\rho(t_\mathrm{MC})$ is just one sigma away
from zero. Above $t_\mathrm{W}$ we assume a single exponential decay with a decay rate
given by the extrapolated value from previous simulations
$\tau_\mathrm{exp}\approx 14 \times t_0/a^2$, which corresponds to
$\tau_\mathrm{exp}\approx 0.2$k MDU for this ensemble. Its amplitude is fixed
by the central value of $\rho(t_\mathrm{W})$.

Note that the original procedure proposed in~\cite{Schaefer:2010hu} uses a
smaller value of $t_\mathrm{W}$ which generally leads to larger autocorrelation times.
In this sense, the  procedure adopted here is less conservative than what is
denoted ``upper bound'' in~\cite{Schaefer:2010hu}, but more conservative than
simply ignoring the tail.  The hereby measured integrated autocorrelation times
are compatible with the estimate of the extrapolated $\tau_\mathrm{exp}$, as
expected for these slow quantities, which serves as a self-consistency check.
With an estimated $\tau_\mathrm{exp}\approx 200$\,MDU, our particular
run has a total statistics of about 30 $\tau_\mathrm{exp}$, which is not what
we would ideally like to have, but it is  still (marginally) acceptable.

\begin{table*}  \small\footnotesize
  \caption{\label{t:t0lmax1}
           Data of hadronic and reference scales and their ratio including
           continuum extrapolations. Set \texttt{A} refers to the 
           $\tilde g_0^2$ values of the CLS simulations and set \texttt{B} 
           to the integer values of $L/a = (a\mu_{\rm had})^{-1}$ of the 
           small-volume simulations.
          }
  \begin{ruledtabular}
  \begin{tabular}{lllllllll}
    &\multicolumn{4}{c}{$\mu_{\rm had,1}$} & \multicolumn{4}{c}{$\mu_{\rm had,2}$}\\
    \cmidrule{2-5}\cmidrule{6-9}
    set&$6/\tilde g_0^2$&$a\mu_{\rm ref}^\star$&$(a\mu_{\rm had})^{-1}$&$\mu_{\rm ref}^\star/\mu_{\rm had}$&$6/\tilde g_0^2$&$a\mu_{\rm ref}^\star$&$(a\mu_{\rm had})^{-1}$&$\mu_{\rm ref}^\star/\mu_{\rm had}$\\
    \midrule
    \multirow{6}{*}{\texttt{A}} &&&&&&&\\
     & $3.3985$  & $0.20899(20)$  & $11.974(79)$   & $2.503(17)$ & $3.3985$  & $0.20899(20)$  & $11.017(68)$  & $2.303(14)$ \\ 
     & $3.4587$  & $0.18476(33)$  & $13.524(46)$   & $2.499(10)$ & $3.4587$  & $0.18476(33)$  & $12.431(39)$  & $2.297(08)$ \\ 
     & $3.549 $  & $0.15556(26)$  & $15.971(34)$   & $2.484(07)$ & $3.549 $  & $0.15556(26)$  & $14.688(49)$  & $2.285(08)$ \\ 
     & $3.6992$  & $0.12059(22)$  & $20.460(60)$   & $2.467(08)$ & $3.6992$  & $0.12059(22)$  & $18.869(48)$  & $2.275(07)$ \\ 
     & $3.8494$  & $0.09437(19)$  & $25.875(109)$  & $2.442(11)$ & $3.8494$  & $0.09437(19)$  & $23.846(93)$  & $2.250(10)$ \\ 
     & \multicolumn{3}{l}{\textbf{Continuum extrapolations:}} &
     & \multicolumn{3}{l}{\textbf{Continuum extrapolations:}}\\
     & \multicolumn{3}{c}{with  $(a\mu_{\rm had})^2<0.07$} & $2.426(18)$ & 
       \multicolumn{3}{c}{with  $(a\mu_{\rm had})^2<0.07$} & $2.240(16)$ \\                                              
     & \multicolumn{3}{c}{with  $(a\mu_{\rm had})^2<0.10$} & $2.433(15)$ & 
       \multicolumn{3}{c}{with  $(a\mu_{\rm had})^2<0.10$} & $2.247(11)$  \\                                              
    \multirow{5}{*}{\texttt{B}} &&&&&&&\\                     
     & $3.3998$  & $0.20845(19)$  & $12.000(58)$  & $2.501(12)$ & $3.4407$ & $0.19137(19)$  & $12.000(34)$  & $2.296(07)$ \\
     & $3.5498$  & $0.15544(21)$  & $16.000(30)$  & $2.487(06)$ & $3.5979$ & $0.14282(19)$  & $16.000(50)$  & $2.285(08)$ \\
     & $3.6867$  & $0.12304(21)$  & $20.000(83)$  & $2.461(11)$ & $3.7372$ & $0.11332(21)$  & $20.000(64)$  & $2.266(08)$ \\
     & $3.8000$  & $0.10234(17)$  & $24.000(105)$ & $2.456(11)$ & $3.8521$ & $0.09396(19)$  & $24.000(109)$ & $2.255(11)$ \\
     & $3.9791$  &   & $32.000(153)$ &  & $4.0336$ &   & $32.000(155)$ \\
     & \multicolumn{3}{l}{\textbf{Continuum extrapolations:}} &
     & \multicolumn{3}{l}{\textbf{Continuum extrapolations:}}\\
     & \multicolumn{3}{c}{with  $(a\mu_{\rm had})^2<0.07$} & $\bf 2.428(18)$ & 
       \multicolumn{3}{c}{with  $(a\mu_{\rm had})^2<0.07$} & $\bf 2.233(17)$ \\                                              
   \end{tabular}
    \end{ruledtabular}
    \label{t:t0lmax1}
\end{table*}

We finally mention that the simulation parameters yielded $\phi_4=1.108(7)$ and
we thus had to perform only a very small shift in the quark masses to reach $\phi_4=1.11$, which is
already accounted for in Table~\ref{t:t0lmax1}.

\section{(III) Determination of $\muref^\star/\muhad$}

\subsection{Interpolation of $a\mu_{\rm had}$ and  $a\mu_{\rm ref}^\star$}

The pairs $(\tilde g_0,a\mu_{\rm had})$ and $(\tilde g_0, a\mu_{\rm
ref}^\star)$ are not yet known at the same values of $\tilde g_0$.  In order to
obtain both $a\mu_{\rm ref}^\star$ and $a\mu_{\rm had}$ at equal values of the
lattice spacing $a$, we have two possibilities: either we fit the values of
$a\mu_{\rm had}$ as a function of $\tilde g_0$  and interpolate to the values
of $\tilde g_0$ where $a\mu_{\rm ref}^\star$ is known, or we interpolate
$a\mu_{\rm ref}^\star$ as a function of $\tilde g_0$  and determine its values
at those $\tilde g_0$ where $a\mu_{\rm had}$       is known.  These procedures
have been labelled sets \texttt{A} and \texttt{B}, respectively in
Table~\ref{t:t0lmax1}.

For the case of set \texttt{A}, $\log(a\mu_{\rm had})$  is fitted to a
polynomial form 
\begin{equation}\label{eq:intrplt_A}
  \log(a\mu_{\rm had}) = \sum_{n=0}^3 c_n (6/\tilde g_0^2)^{n} \,,
\end{equation}
which allows to determine values of $a\mu_{\rm had}$ at the $\tilde g_0^2$
where $a\muref^\star$ is known (set \texttt{A} of Table~\ref{t:t0lmax1}). We
have repeated the whole analysis chain by replacing this interpolation by the
purely heuristic (inverse) function $ (6/\tilde g_0^2)=\sum_{n=0}^3 p_n
(a\mu_{\rm had})^{-n}$, with entirely compatible results.

\begin{figure*}
  \includegraphics[width=0.85\textwidth]{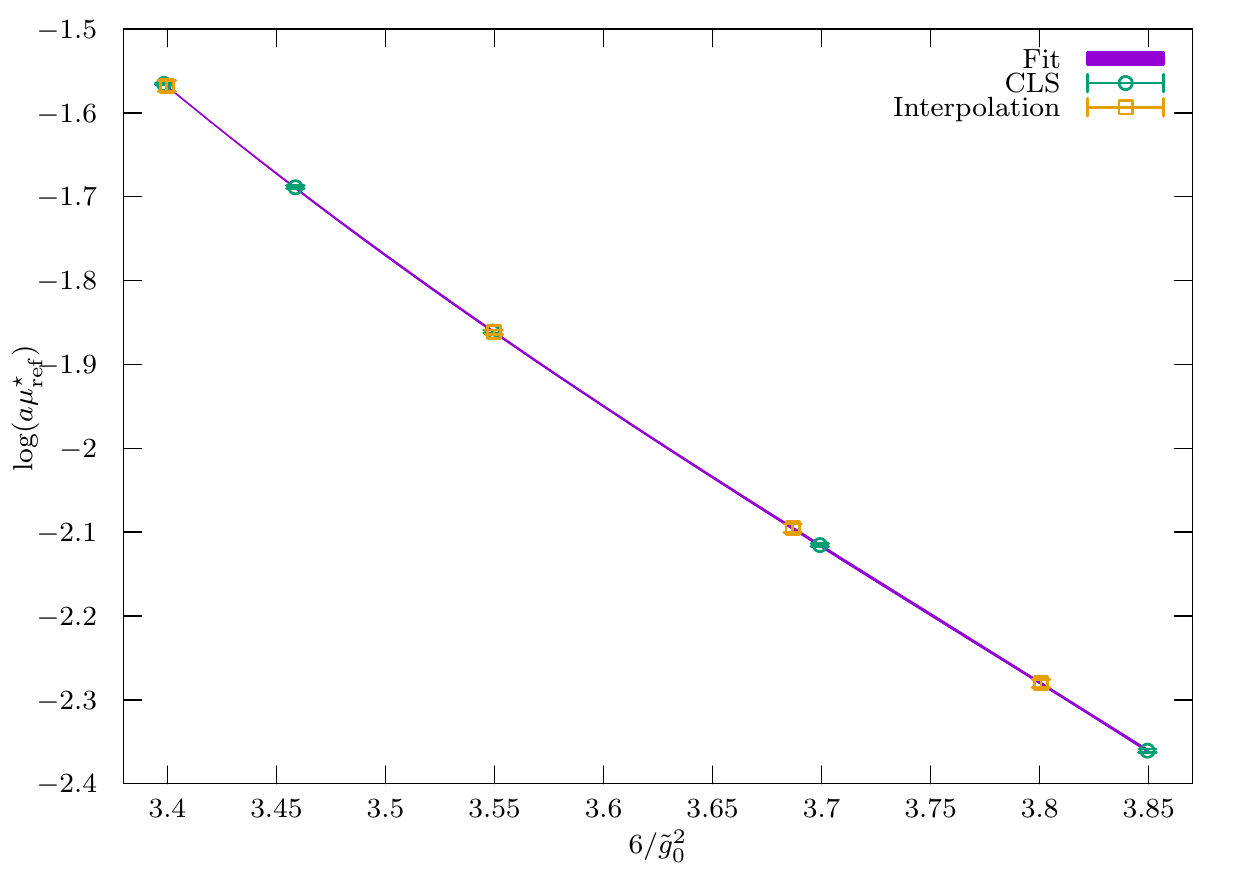}
  \caption{\label{fig:fitt0}
           Fit of $\log(a\mu_{\rm ref}^\star)$  to the points marked CLS. This
           fit allows us to read the values of $a\mu_{\rm ref}^\star$ at the
           values of $g_0$ where $a\mu_{\rm had,1}$ is known, set {\tt B}  (points labelled
           Interpolation).
          }
\end{figure*}

Regarding procedure {\tt B}, the desired $a\mu_{\rm ref}^\star\equiv
a/\sqrt{8t^{\scalebox{0.6}{\rm sym}}_0}$ as a function of $\tilde g_0$ is found
in a very similar way. The functional form
\begin{equation} \label{eq:fit_t0}
   \log(t_0^{\scalebox{0.6}{\rm sym}}/a^2) = \sum_{n=0}^3 a_n \,\left(\frac{6}{\tilde g_0^2}-3.5\right)^{n} \,,  \end{equation}
describes our data very well ($\chi^2/{\rm dof} = 0.5 / 1$) and is generally
useful for scale setting. Its coefficients $a_n$ with covariances are listed in
Table~\ref{tab:an_cov}.  Figure~\ref{fig:fitt0} shows the fit to the original
CLS data and the interpolated values of $\log(a\mu_{\rm ref}^\star)$
at the values of $\tilde g_0$ determined from our data. {\cred }

\begin{table*}[ht]
  \footnotesize
  \caption{\label{tab:an_cov}
           Parameters $a_n$ and covariances ${\rm cov}(a_n,a_m)$ of
           our preferred parametrization, eq.~\eqref{eq:fit_t0}.
          }
  \begin{ruledtabular}
                \begin{tabular}{crrrrr}
     $n$ & $a_n\quad$     & ${\rm cov}(a_n,a_0)$ & ${\rm cov}(a_n,a_1)$ &
     ${\rm cov}(a_n,a_2)$ & ${\rm cov}(a_n,a_3)$\\ 
     \midrule
 $0$   & $1.45992$ & $ 0.00001$ & $   0.00001$ & $  -0.00055$ & $   0.00129$ \\
 $1$ & $3.78480$ & $  0.00001$ & $   0.00033$ & $  -0.00036$ & $  -0.00225$ \\
 $2$ & $-2.10252$ & $  -0.00055$ & $  -0.00036$ & $   0.05423$ & $  -0.14131$ \\
 $3$ & $2.71565$ & $  0.00129$ & $  -0.00225$ & $  -0.14131$ & $   0.40952$ \\
  \end{tabular}
\end{ruledtabular}
\end{table*}

\subsection{Continuum extrapolation of $\muref^\star/\muhad$ }

The dimensionless ratio $\muref^\star/\muhadi{}$ plotted in
Figure~4 in the main text and listed in Table~\ref{t:t0lmax1}
can now be extrapolated to the continuum.  We fit the data linearly in $a^2$,
dropping either points above $(a\muref^\star)^2 =0.07$  or above
$(a\muref^\star)^2 =0.1$.  We performed this analysis with both sets \texttt{A}
and \texttt{B}.  The values at $a=0$ are fully compatible, see Table~\ref{t:t0lmax1} and
Figure~4 of the main text. We also extrapolated $\log(\muref^\star/\muhadi{})$ and
$(\muref^\star/\muhadi{})^{-1}$, which of course have different higher order
discretisation errors. Continuum values do not change by more than a per-mille.
One may also combine sets \texttt{A} and \texttt{B} either after the
extrapolation or before. That variant yields somewhat smaller errors and fully
compatible central values in the continuum.

As our final estimate we take the numbers of set \texttt{B} extrapolated with
data satisfying $(a\muref^\star)^2 \leq 0.07$ since these have the larger
errors and in particular, since these make optimal use of our CLS point closest
to the continuum at $a\approx 0.039\,\fm$.  That data point is in any case crucial in
stabilizing  the continuum extrapolation. It renders the difference between the
last data point and the extrapolated value insignificant and allows us to cite
a continuum ratio with less than a percent error.

\bibliography{gbar-nf3-final}

\end{document}